\documentclass[acmtog,authorversion]{acmart}

\usepackage{booktabs} 
\usepackage[ruled]{algorithm2e} 

\SetAlFnt{\small}
\SetAlCapFnt{\small}
\SetAlCapNameFnt{\small}
\SetAlCapHSkip{0pt}

\usepackage{color}
\definecolor{olive}{rgb}{0.1,0.8,0.3}
\definecolor{mauve}{rgb}{0.48,0,0.72}

\acmJournal{TOG}
\acmVolume{43}
\acmNumber{1}

\setcopyright{none}

\acmDOI{10.1145/3625264}

\begin{document}

\title[SparsePoser: Real-time Full-body Motion Reconstruction from Sparse Data]{SparsePoser: Real-time Full-body Motion Reconstruction \\ from Sparse Data}


\author{Jose Luis Ponton}
\authornote{Jose Luis Ponton and Haoran Yun are equal contributors to this work and designated as co-first authors.}
\orcid{0000-0001-6576-4528}
\affiliation{%
  \institution{Universitat Politècnica de Catalunya}
  \city{Barcelona}
  \country{Spain}
  \postcode{08034}
}
\email{jose.luis.ponton@upc.edu}

\author{Haoran Yun}
\authornotemark[1]
\orcid{0000-0001-6192-6673}
\affiliation{%
  \institution{Universitat Politècnica de Catalunya}
  \city{Barcelona}
  \country{Spain}
  \postcode{08034}
}
\email{haoran.yun@upc.edu}

\author{Andreas Aristidou}
\orcid{0000-0001-7754-0791}
\affiliation{%
  \institution{University of Cyprus}
  \streetaddress{75, Kallipoleos}
  \city{Nicosia}
  \country{Cyprus}
  \postcode{1678}
}
\affiliation{%
  \institution{CYENS Centre of Excellence}
  \streetaddress{23, Plateia Dimarchias}
  \city{Nicosia}
  \country{Cyprus}
  \postcode{1016}
}
\email{a.aristidou@ieee.org}

\author{Carlos Andujar}
\orcid{0000-0002-8480-4713}
\affiliation{%
  \institution{Universitat Politècnica de Catalunya}
  \city{Barcelona}
  \country{Spain}
  \postcode{08034}
}
\email{andujar@cs.upc.edu}

\author{Nuria Pelechano}
\orcid{0000-0002-1437-245X}
\affiliation{%
  \institution{Universitat Politècnica de Catalunya}
  \city{Barcelona}
  \country{Spain}
  \postcode{08034}
}
\email{npelechano@cs.upc.edu}

\renewcommand\shortauthors{Ponton~et al.}

\begin{abstract}

Accurate and reliable human motion reconstruction is crucial for creating natural interactions of full-body avatars in Virtual Reality (VR) and entertainment applications. As the Metaverse and social applications gain popularity, users are seeking cost-effective solutions to create full-body animations that are comparable in quality to those produced by commercial motion capture systems. In order to provide affordable solutions though, it is important to minimize the number of sensors attached to the subject's body. Unfortunately, reconstructing the full-body pose from sparse data is a heavily under-determined problem. Some studies that use IMU sensors face challenges in reconstructing the pose due to positional drift and ambiguity of the poses. In recent years, some mainstream VR systems have released 6-degree-of-freedom (6-DoF) tracking devices providing positional and rotational information. Nevertheless, most solutions for reconstructing full-body poses rely on traditional inverse kinematics (IK) solutions, which often produce non-continuous and unnatural poses. In this paper, we introduce SparsePoser, a novel deep learning-based solution for reconstructing a full-body pose from a reduced set of six tracking devices. Our system incorporates a convolutional-based autoencoder that synthesizes high-quality continuous human poses by learning the human motion manifold from motion capture data. Then, we employ a learned IK component, made of multiple lightweight feed-forward neural networks, to adjust the hands and feet towards the corresponding trackers. We extensively evaluate our method on publicly available motion capture datasets and with real-time live demos. We show that our method outperforms state-of-the-art techniques using IMU sensors or 6-DoF tracking devices, and can be used for users with different body dimensions and proportions. 
\end{abstract}

%
%


%
%


\begin{teaserfigure}
\centering
 \includegraphics[width=1\linewidth]{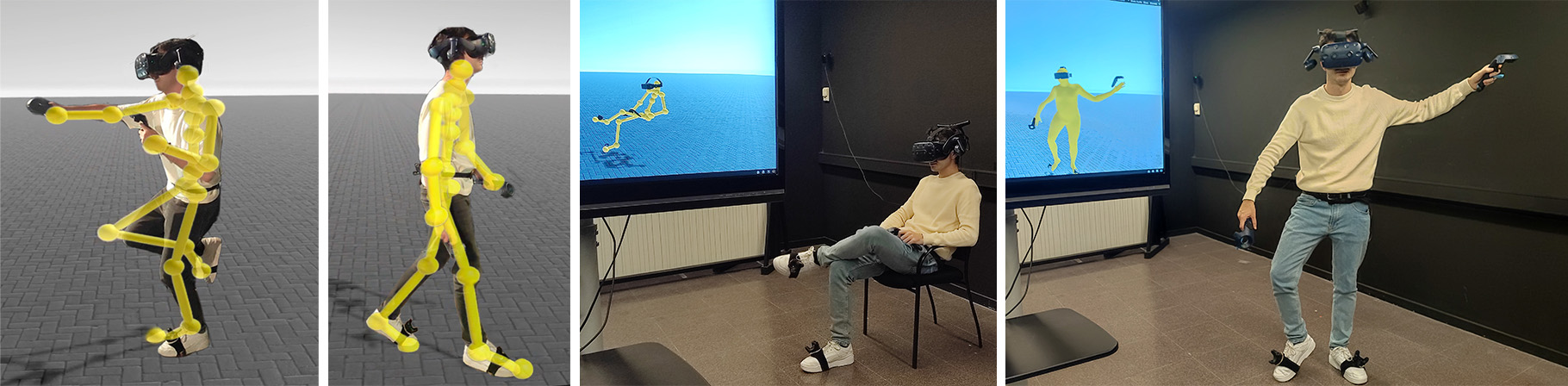}
 \caption{Highly accurate poses reconstructed from six 6-DoF trackers. On the left, a virtual reconstructed skeleton is rendered over the user. On the right, the virtual skeleton on the screen mimics the user poses and has its end-effectors in the correct position.}
\label{fig:teaser}
\end{teaserfigure}

\maketitle

\section{Introduction}

Real-time human motion reconstruction is essential in many Virtual Reality (VR) and Augmented Reality (AR) applications in areas such as entertainment, simulation, training, sports and education. 
With the growing interest in having users collaborate in the Metaverse and social applications, the need to have avatars that represent those users is rapidly increasing. Having our own virtual avatar can provide embodiment, but also seeing avatars representing other users can enhance non-verbal communication and the overall sense of presence. Therefore, it is essential to have high-quality animations for avatars that can convey our movements accurately.

Currently, most of the established technologies for high-quality motion capture (mocap), such as Vicon or Xsens, use optical markers or inertial measurement units (IMUs). Unfortunately, these systems are cost-demanding, require specialized personnel, and need extended and complex calibration processes, thus, are most suitable for large companies or research labs. As VR and AR technologies become increasingly affordable to the general public, there needs to be a similar trend for mocap systems. Ideally, consumer-grade VR and AR should also offer affordable and accurate mocap, with easy configuration and installation, to serve home users who want inexpensive but reliable means for full-body interaction with applications for VR/AR, social interaction and entertainment.

Given the small number of input trackers in consumer-grade VR/AR, most current solutions are limited to representing the user's upper body animated with Inverse Kinematics (IK) by employing the Head-Mounted Display (HMD) and hand-held controllers as end-effectors. The lack of full-body tracking, though, breaks the immersion when users look at themselves or collaborate with other participants~\citep{Fribourg:2020, Debarba:2020, Toothman:2019}. Recent works address this issue by predicting the full-body pose from three 6-DoF tracking devices (HMD + two hand-held controllers) \citep{Winkler:2022, Ponton:2022b, Jiang:2022b}. However, the absence of lower-body tracking makes the problem highly under-determined, thus limiting the lower-body motion to basic locomotion. 

Other works \citep{Huang:2018, Jiang:2022, Yi:2021, Yi:2022} use a sparse set of IMUs (e.g., six) to reduce the cost of motion capture systems such as Xsens (17 IMUs) while still being able to represent a broader range of motion by placing some sensors on the lower body. IMU-based approaches have become increasingly popular due to their advantages in certain applications. One notable advantage is that they do not require external devices, thereby allowing for a capture volume that is not restricted by physical limitations or environmental factors such as illumination or occlusions. Unfortunately, standalone IMUs for full-body mocap are not included as standard options in mainstream VR/AR systems. In addition, they require careful initialization, rely on previous pose predictions and suffer from positional drift. Incorrect pose prediction happens because IMUs may produce the same sensor output for very distinct poses (e.g., standing still and sitting down), and thus, depend on the previous pose being correct to compute the next pose. In the case of VR, the positional drift problem can be very noticeable when the self-avatar position moves away from the user or when accurate end-effector placement is needed.

Some VR/AR systems can be used with additional rotational and/or positional tracking devices, such as HTC VIVE Trackers. One limitation of this technique is their reliance on external devices. The most common solution to include full-body avatars is to apply IK using these devices (along with the HMD and hand-held controllers) as end-effectors \citep{Oliva:2022, Ponton:2022}. Having absolute positional information allows these methods to improve pose synthesis and to place the avatar and its end-effectors accurately. The drawback is that IK usually generates static, unnatural, and time-incoherent human motion, often leading to popping artifacts in some joints, such as the knees.

In this paper, we present SparsePoser, a novel data-driven method for animating avatars using only six tracking devices with 6-DoF (see Figure~\ref{fig:teaser}). SparsePoser works by encoding the information retrieved from the sensors and the static representation of the user (i.e., the skeleton), and decoding it to a full-body pose by reconstructing all joints between the end-effectors. We also introduce a learned IK step that can accurately re-position the end-effectors according to the sensor's information. The main contributions of our paper are:
\begin{itemize}
    \item To the best of our knowledge, SparsePoser is the first deep learning-based system to reconstruct full-body motion from a sparse set of positional and rotational sensors such as those found in recent consumer-grade VR/AR systems. Unlike approaches using three sensors to reconstruct the upper-body pose and roughly guess the lower-body pose, SparsePoser accurately recovers motion for the whole body.
    \item A deep learning-based architecture to synthesize human motion in real time consisting of: (a) a generator, which is a convolutional-based autoencoder using skeletal-aware operations, inspired by \citet{Aberman:2020}, that learns the human motion features from sparse input and produces highly smooth and realistic poses; and (b) a learned IK network that adjusts the limbs of the human skeleton towards the end-effectors' positions and rotations.
    \item A VR-specific motion capture database created from hours of users interacting and navigating in VR applications wearing a HMD and Xsens. This is the first database that gathers the kind of interaction movements and locomotion that are most relevant to VR avatar animation.
\end{itemize}

We showcase the effectiveness of SparsePoser by comparing it to state-of-the-art techniques that use IMU sensors or 6-DoF trackers. The evaluation consists of a quantitative analysis using publicly available datasets, and a qualitative analysis through real-time demonstrations.
Furthermore, we assess the various components of our system, including the chosen pose representation, and examine its ability to adapt to users of varying heights and body proportions.
\section{Related Work}
\label{Section:Related_Work}
The literature on human motion reconstruction is extensive and encompasses a wide range of research. This section briefly reviews methods that utilize sparse sensor signals from IMUs and VR tracking devices. We first discuss the general problem of full-body reconstruction from low-dimensional input and, subsequently, delve into the specific problem of learning-based IK methods.

\subsection{Full-body Motion Reconstruction from Sparse Input}
Using a reduced set of IMUs placed on a user's body to reconstruct human motion has been extensively investigated in past years. One of the advantages of IMUs is that they do not require external sensors or cameras and thus, can be used in any lighting condition or environment, and do not suffer from occlusion problems. Compared to commercial IMU-based motion capture suits \citep{Xsens}, recent methods are less intrusive and easier to set up due to the use of a lower number of sensors.

Early work on kinematic models with six IMUs, e.g., \citet{Marcard:2017}, propose an optimization-based offline method that reconstructs full-body poses. Further works, such as DIP \citep{Huang:2018} and TransPose \citep{Yi:2021}, use deep learning-based models, like recurrent neural networks (RNN), that learn from large motion capture datasets and can reconstruct poses in real time with higher accuracy. As IMUs provide no positional information, it is challenging to correctly estimate the global positions or translations of the user in the virtual environment. DIP concentrated on the generation of poses while fixing the character's position. TransPose uses RNN with a supporting-foot-based method to predict global translations. Transformer-based models \citep{Vaswani:2017}, initially proposed for natural language processing, have been extensively used in many domains with sequential inputs. In that manner, \citet{Jiang:2022} introduce a conditional Transformer decoder model that reconstructs full-body pose and can correct the drift by predicting stationary body points with soft-IK constraints, stabilizing the generated root velocity and joint angles. Apart from kinematic models, physics-based methods have also been used for motion reconstruction with IMUs. \citet{Yi:2022} propose a framework that combines an RNN-based kinematic module with a physics-based optimizer to generate physically plausible motions from a sparse set of IMUs. 

Overall, one significant drawback of IMU-based techniques is that after prolonged usage, the rotational and translation information reconstructed tends to drift due to the double integration needed to retrieve positions from accelerations. This issue leads to inaccurate global translation of the character and accumulated positional and rotational errors on the body pose. In order to minimize drift, recent work by \citet{yi:2023} leverages the use of a monocular camera to locate the human within the reconstructed scene through simultaneous localization and mapping (SLAM). We opt to use commercial VR hardware that combines both IMUs and photosensors, to provide precise position and orientation of each tracked object while not suffering from drift over time. 

As commercial VR devices become widely available, some works reconstruct full-body poses from only the Head-Mounted Display (HMD) and two hand-held controllers. Data-driven methods have proven to be able to reconstruct high-quality and continuous poses for certain applications. \citet{Dittadi:2021} use a variational autoencoder to reconstruct full-body poses from three-tracking points, but without estimating global translations. \citet{Winkler:2022} propose a reinforcement learning framework that, together with a physics simulator, generates natural and physically plausible movements. \citet{Jiang:2022b} present a Transformer-based encoder to estimate the full-body poses and global rotations in real time. \citet{Aliakbarian:2022} harness the advantages of generative models to introduce a conditional flow-based model capable of generating plausible full-body poses from sparse input. Other methods \citep{Ahuja:2021, Ponton:2022b} explore the idea of searching in a motion dataset, similar to Motion Matching \citep{Clavet:2016}, to find a sequence of full-body poses that match the current pose and user input, hence ensuring the quality of the motion. However, using only three tracking points provides limited full-body information, especially for the lower body where almost no information can be recovered, resulting in motion with foot-sliding problems, and contact-point violations. Therefore, these methods can only be used in certain applications with limited lower-body motion, such as locomotion.

Another common approach is to add additional trackers to the user to reduce ambiguity. With one additional tracker on the user's pelvis, \citet{Yang:2021} propose an RNN-based model with Gated Recurrent Units (GRUs) that utilizes velocity data to accurately predict low-body movements, including global translation and orientation. Nonetheless, their upper-body poses are solved by an IK solver, thus, providing lower-quality upper-body poses. Adding trackers on the pelvis and feet, IK solvers are being explored to generate full-body poses \citep{Zeng:2022, FinalIK, Ponton:2022, Oliva:2022}. However, as these methods mostly optimize the pose to reach the end-effectors, the generated motion sequences may lack temporal coherency and produce unnatural non-human-like poses. When used in VR, such problems can negatively impact the Sense of Embodiment \citep{Fribourg:2020, Goncalves:2022}. To overcome these issues, our method uses a two-stage approach that combines a convolutional-based model with skeleton-aware operations and a learned IK model, achieving smooth high-quality poses while maximizing the end-effector accuracy.

\subsection{Learned Inverse Kinematics}

In robotics and computer animation, it is common to enforce an Inverse Kinematics (IK) solver to determine the positions and orientations of the intermediate joints in a kinematic chain when the positions and orientations of the end-effectors (leaf joints) are known.
\citet{Aristidou:2018} comprehensively reviews the most popular IK approaches for reconstructing human motion, such as analytical and numerical IK solvers. Furthermore, a combination of IK solvers can be utilized to solve the pose of a human-like character from the end-effectors. For instance, RootMotion's Final IK \citep{FinalIK} uses a combination of analytical and heuristic solvers to solve the pose of different body parts.
However, traditional IK solvers typically present scalability limitations for multi-chain characters, and a trade-off between computational efficiency and naturalness of the generated poses, as noted by \citet{Caserman:2019}.

Traditional IK solvers are primarily focused on optimizing the alignment of end-effectors with their corresponding leaf joints. However, they often struggle to generate natural human-like poses. To overcome this limitation, some studies combine data-driven methods, which can learn poses from high-quality motion capture data, with traditional IK to achieve accurate end-effector placement.
For example, \citet{Jiang:2022b} employ an IK module in their Transformer-based pipeline, to adjust the shoulder and elbow positions, and to avoid deviations between the predicted hand positions and the tracked VR controllers. \citet{Ponton:2022b} use an IK algorithm to solve the pose of the arms after a Motion Matching module produces a full-body pose. Similarly, \citet{Yang:2021} uses a deep learning-based method for solving the lower body and an IK solver for the upper body. While these solutions represent a good compromise between pose quality and end-effector accuracy, incorporating the last IK step may override the pose generated by the data-driven solution, and thus, it may incorporate all the issues typically found in traditional IK solutions.

As motion data becomes more widely available, data-driven IK solutions have consistently attracted attention in robotics and computer animation.
In robotics, learned IK methods employ neural networks, such as light-weighted feed-forward networks \citep{Bocsi:2011, Duka:2014, Csiszar:2017, Bensadoun:2022}, generative adversarial networks \citep{Ren:2020} and conditional normalizing flow networks \citep{Ames:2022}, to learn a fixed solution or explore the space of possible solutions for a given target end-effector. These methods accelerate the IK computation but are limited to specific kinematic chains; typically robotic arms with a low number of degrees of freedom when compared with a human body. 

In computer animation, previous work focus on using machine learning models for reconstructing full-body poses. \citet{Grochow:2004} and \citet{Wu:2011} present an IK system based on scaled Gaussian processes to model a probability distribution over the space of poses, and use different training data to generate various styles. \citet{Huang:2017} utilizes a multi-variate Gaussian model as soft constraints for a Jacobian-based IK solver to obtain a sequence of coherent nature poses in real time. All these methods can generate natural poses, but at the expense of being less efficient than conventional IK solvers. Moreover, the use of Gaussian processes severely limits the size of the training set, and thus, the method fails to generate natural poses when the desired pose deviates significantly from the training poses. In our work, we utilize the recent advancements in deep learning-based models, which can be trained with a large number of poses, to overcome these limitations.

Recently,~\citet{Victor:2021} introduce an IK solver that is based on an autoencoder structure, which aligns hand joints to the target position from a starting pose. However, their model has some limitations as it modifies the entire skeleton each time an end-effector is altered, resulting in the emergence of foot-sliding artifacts and a lack of temporal coherence. Furthermore, it only predicts joint positions, making it skeleton-dependent and lowering the skeletal degrees of freedom. In contrast, the goal of our learned IK component is to enhance the high-quality pose synthesized by the generator by leveraging the strengths of our convolutional-based generator, while, at the same time, addressing the issues of foot-sliding and increasing end-effector accuracy. 

\citet{Zhou:2020} introduced the network IKNet consisting of one fully-connected network that computes joint rotations from joint positions and bone orientations of the hand. Their approach does not learn to modify the pose; instead, it converts a hand pose provided by the joint positions into joint rotations. In contrast, our learned IK, given an initial body pose and target end-effectors (hands and feet), modifies each limb independently to better reach the targets. In addition, it learns to deal with the more complex articulation of full bodies.
\section{Overview}
\label{Section:Overview}

\begin{figure*}[ht]
  \includegraphics[width=1\linewidth]{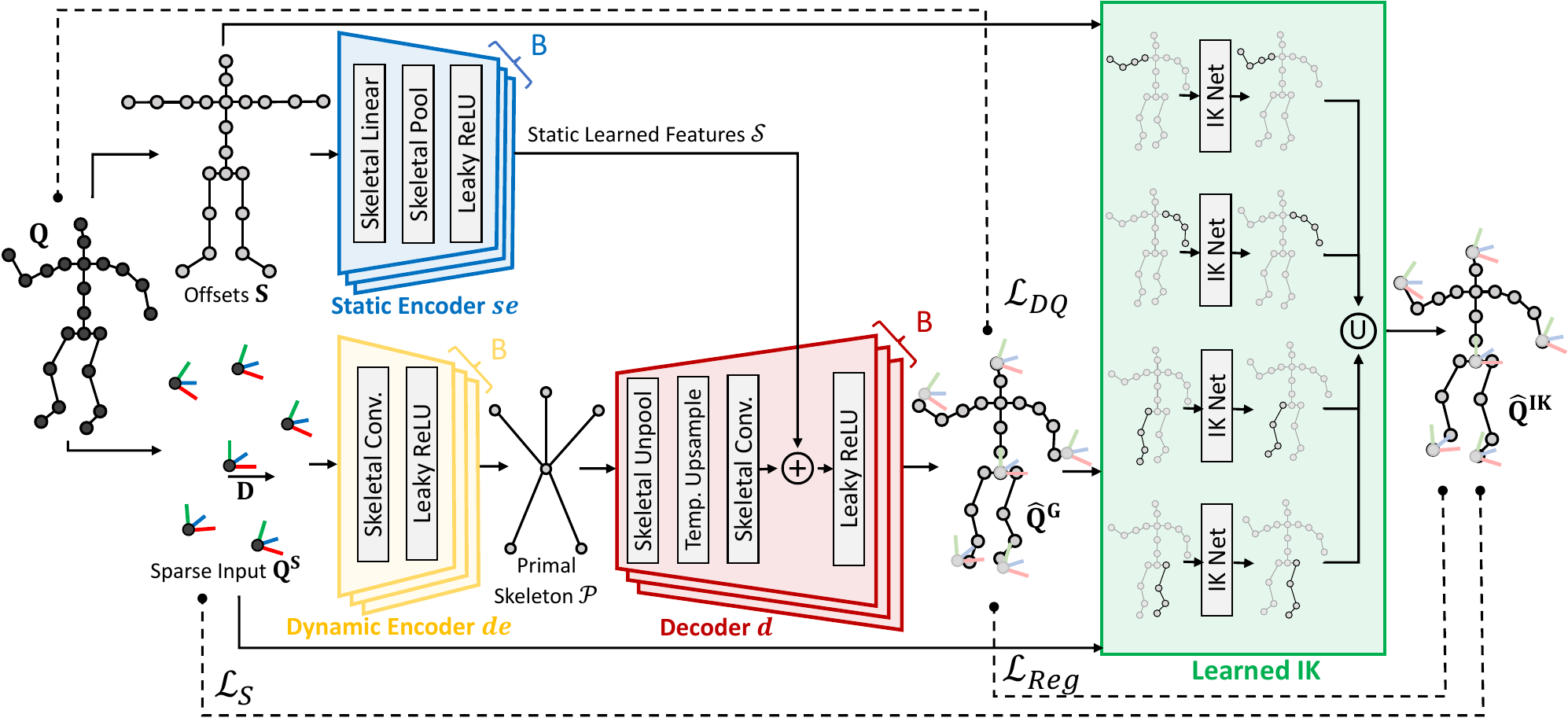}
  \caption{Network architecture of SparsePoser for reconstructing full-body pose from sparse data. First, the static structure of the skeleton $\mathbf{S}$, the sparse input $\mathbf{Q^S}$, and the displacement $\mathbf{D}$ are extracted from the motion $\mathbf{Q}$. A convolutional-based autoencoder (composed of the Static and Dynamic Encoders, $se$ and $de$, and the Decoder, $d$) learns to reconstruct user poses for a set of contiguous frames. Subsequently, a learned IK solver adjusts the positions of the end-effectors to attain the target positions and rotations.}
  \label{fig:pipeline}
\end{figure*}

This paper presents a deep learning-based framework for animating human avatars from a sparse set of input sensors. A visual diagram of SparsePoser is shown in Figure~\ref{fig:pipeline}. First, we retrieve the positions and rotations from six sensors placed on the head, hands, feet and pelvis (the root in our case) of the user. Then, these are transformed into a root-centered dual quaternion-based pose representation~\citep{Andreou:2022}, which allows the network to implicitly understand the structure of the skeleton and synthesize accurate poses. A convolutional-based autoencoder extracts the main features from the sensors and reconstructs the user poses for a set of contiguous frames. 
This initial stage utilizes skeleton-aware operations, similar to~\citep{Aberman:2020}, to maintain consistency and generate accurate human postures. Subsequently, we integrate a learned IK solver that has been trained to adjust the positions of the end-effectors to attain the targeted points.
Once trained, our method can be applied to different-sized users using standard commercial VR systems that provide rotational and positional information, such as HTC VIVE Trackers.

\section{Background}
\label{Section:Background}

This section provides the fundamental concepts essential to understanding the proposed method. Specifically, we introduce dual quaternions, which serve as the pose representation utilized in Section~\ref{Section:InputAndPose} as presented by \citet{Andreou:2022}, as well as the skeleton-aware operations introduced by \citet{Aberman:2020} which we use as a component of our network architecture.

\paragraph{Dual Quaternions}
A dual quaternion $\mathbf{\underline{q}} \in \mathbb{R}^{8}$ can be represented as two quaternions $\mathbf{q_r} \in \mathbb{R}^4$ and $\mathbf{q_d} \in \mathbb{R}^4$ in the form $\mathbf{\underline{q}} = \mathbf{q_r} + \epsilon \mathbf{q_d}$, where $\mathbf{q_r}$ and $\mathbf{q_d}$ are the real and dual part, respectively, and $\epsilon$ is the dual unit. A dual quaternion $\mathbf{\underline{q}}$ is unit if $\mathbf{\underline{q}} \otimes \mathbf{\underline{q}}^{*} = 1$, where $\mathbf{\underline{q}}^{*}$ its the conjugate of $\mathbf{\underline{q}}$. 

Let $\mathbf{q_r} = \cos{\frac{\theta}{2}} + \mathbf{\hat{u}} \sin{\frac{\theta}{2}}$ be a quaternion representing a rotation $\theta$ about the unit vector $\mathbf{\hat{u}}$, and $\mathbf{t} = (t_1, t_2, t_3)$ be a translation and its corresponding pure quaternion $\mathbf{q_t} = 0 + \mathbf{t}$. We can compactly represent a rigid displacement \citep{Jia:2013, Kavan:2007} with a unit dual quaternion as follows:
\begin{align}
    \mathbf{\underline{q}} &= \mathbf{q_r} + \frac{\epsilon}{2} \mathbf{q_t} \otimes \mathbf{q_r} \\
    &= \cos{\frac{\theta}{2}} + \mathbf{\hat{u}} \sin{\frac{\theta}{2}} + \frac{\epsilon}{2} \left( -\sin{\frac{\theta}{2}} (\mathbf{t} \cdot \mathbf{\hat{u}}) + \cos{\frac{\theta}{2}}\mathbf{t} + \sin{\frac{\theta}{2}} \mathbf{t} \times \mathbf{\hat{u}} \right)
\end{align}
where $\otimes$ denotes the quaternion multiplication. From a unit dual quaternion $\mathbf{\underline{q}} = \mathbf{q_r} + \epsilon \mathbf{q_d}$, we can easily extract the rotation (the quaternion $\mathbf{q_r}$) and the translation 
    $\mathbf{t} = 2 \mathbf{q_d} \otimes \mathbf{q_r}^{*}$

\begin{figure}[htb]
  \includegraphics[width=1\linewidth]{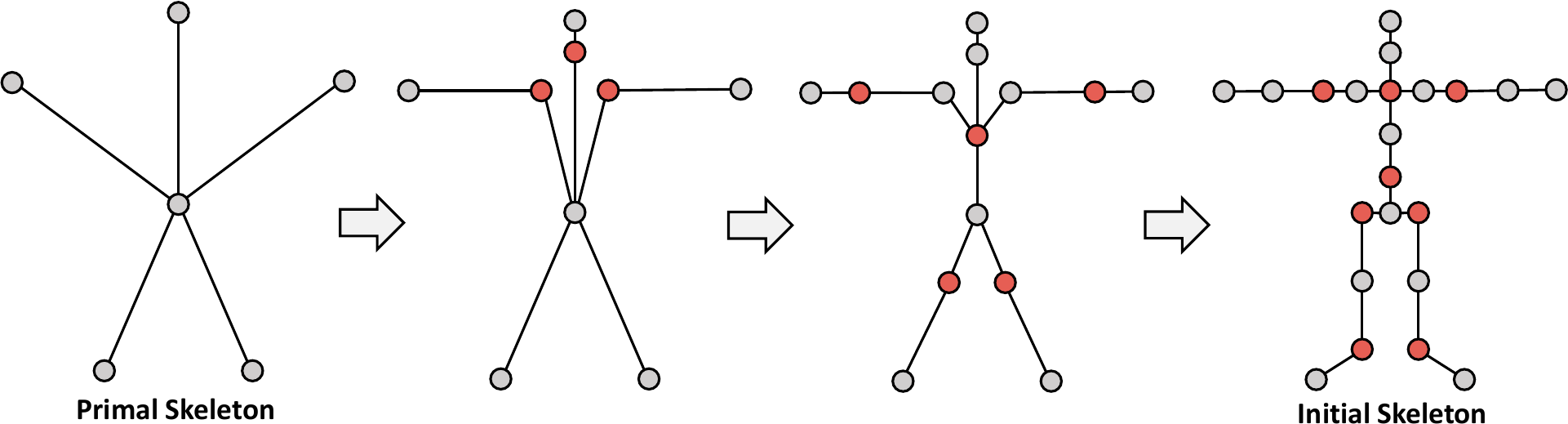}
  \caption{Skeleton unpooling procedure used in the Decoder. A skeleton is pooled by collapsing pairs of consecutive joints, as shown in red, and unpooled by the opposite procedure.}
  \label{fig:skeleton_pooling}
\end{figure}

\paragraph{Skeleton-aware operations}
Both the static and the dynamic autoencoders in Figure~\ref{fig:pipeline} use skeleton-aware operations that explicitly account for the hierarchical bone structure and joint adjacency. Given a skeleton with $J$ joints encoded as a list $\mathrm{J} = (j_0, j_1, \dots, j_J)$, we can represent their hierarchical structure with a list of the same size containing the index of each joint's parent $\mathrm{P} = (p_0, p_1, \dots, p_J)$.
For each joint with index $x$, we also store its neighbors $\mathcal{N}_x = \{j_y \, | \, dist(j_y, j_x) < d , 0 \leq y < J \}$, i.e., the set of joints that, when interpreting the skeleton as a graph, are at a distance less or equal to $d$ (e.g., $d=2$).
A skeleton is pooled by collapsing pairs of consecutive joints until solely leaf and root joints are left, and it is unpooled by the opposite procedure, as shown in Figure~\ref{fig:skeleton_pooling}.
Thus, we have different skeletal structures for each pooling $i$. Suppose we repeat this process $B$ times; we will have $B$ lists of joints $(\mathrm{J}^0, \mathrm{J}^1, \dots, \mathrm{J}^B)$ with their corresponding parents $(\mathrm{P}^0, \mathrm{P}^1, \dots, \mathrm{P}^B)$ and neighbors $(\mathcal{N}^0, \mathcal{N}^1, \dots, \mathcal{N}^B)$.

The skeletal convolution is applied as a standard one-dimensional convolution over the temporal channel at each pooling level $i$, with the difference that the learned weights $\mathrm{\mathbf{W}}^i \in \mathbb{R}^{I \times K \times k}$ ($I$ is the number of input channels, $K$ are the learned filters, and $k$ is the kernel size) are multiplied by a mask $\mathrm{\mathbf{M}}^i \in \mathbb{R}^{I \times K \times k}$ defined as follows:
\begin{equation}
    \mathrm{\mathbf{M}}^i_{x,y} = \begin{cases}
        (1, \dots, 1) \in \mathbb{R}^k \text{ if } j_y \in \mathcal{N}^i_x \\
        (0, \dots, 0) \in \mathbb{R}^k \text{ otherwise }
    \end{cases}
    \label{eq:mask}
\end{equation}

In Equation~\ref{eq:mask}, we assume that each joint is mapped to one channel for simplicity; however, each joint starts with eight channels, as we use dual quaternions for pose representation, and the channels are duplicated after the execution of each block in the Dynamic Encoder, and halved in each block of the Decoder. This allows us to capture higher-level features as the number of joints is reduced. Therefore, the input channels are the number of joints multiplied by the number of channels per joint. The learned filters are similarly defined but use the number of channels per joint of the next block. As seen in Equation~\ref{eq:mask}, when a convolution is performed on a specific joint, the mask only permits neighboring joints to be taken into account.
The skeletal linear operation can be seen as a particular case of the skeletal convolution where $k = 1$.

While our work draws inspiration from the motion retargeting technique proposed by~\citet{Aberman:2020}, our focus is on synthesizing motion from sparse data. Our architecture differs significantly in several key ways. Firstly, our generator is trained to learn the main features of motion from sparse data and reconstruct poses using unpooling operations and simplified loss functions. Secondly, we introduce a novel learned IK network that produces accurate end-effector positioning. Thirdly, we use dual quaternions for pose representation, which significantly improves animation smoothness. Finally, we propose a VR controller that enables our method to be used with VR hardware, making it a more versatile tool for motion synthesis.

\section{Method}
\label{Section:Method}

In this section, we describe the structure of the proposed network for reconstructing full-body poses from a sparse set of trackers. We present the inputs and pose representation, followed by the network structure and the training procedure.

\subsection{Input and Pose Representation}
\label{Section:InputAndPose}

The input of our method is a set of motion sequences of length $T$ (number of poses) using a humanoid skeleton with $J$ joints. We separate it into three components $\mathbf{S}$, $\mathbf{Q}$ and $\mathbf{D}$. 
The static component $\mathbf{S} \in \mathbb{R}^{J \times 3}$, contains a set of offsets (3D vectors) representing the local positions of the joints in the bind pose. Each joint is defined in the local frame of its parent, thus creating a hierarchical skeleton representation. In contrast, the dynamic component $\mathbf{Q} \in \mathbb{R}^{T \times J \times 8}$ contains the root space local rotations and translations of all joints per frame, using dual quaternions as explained below. Finally, the displacement component $\mathbf{D} \in \mathbb{R}^{T \times 3}$ stores the displacement between frames of the root joint, per all frames, as 3D vectors. 

For the dynamic part, $\mathbf{Q}$, we represent the local rotations and translations using unit dual quaternions, as presented by~\citet{Andreou:2022}. 
Dual quaternions provide a unified and compact representation that encodes both rotational and translation information in orthogonal quaternions, allowing the network to understand human motion better. It is also ideal for independently structuring each joint's location and orientation by constructing them relative to the root joint, making our predictions less vulnerable to accumulated errors as we move along the kinematic chain.

\subsection{Network Structure}
The method is structured into two main parts, as represented in Figure~\ref{fig:pipeline}. The first part is the generator, which has the structure of an autoencoder with skeleton-aware operations as building blocks~\citep{Aberman:2020}. The autoencoder learns to reconstruct a full-body pose from a low-dimensional input; it is able to understand the human motion manifold and, thus, produce continuous and highly realistic poses. The second part is a set of neural networks that adjusts the skeleton's limbs toward their corresponding end-effectors.

\paragraph{Generator}
The input of the generator consists of the three components $\mathbf{S}$, $\mathbf{Q}$ and $\mathbf{D}$, which are used to synthesize a full-body pose. It comprises the Static Encoder $se$, the Dynamic Encoder $de$ and the Decoder $d$. Firstly, the Static Encoder, $se$, uses the static component $\mathbf{S}$ to produce a list of $B$ ($B = 3$ in our experiments) static learned features $\mathcal{S} = (\mathcal{S}^0, \mathcal{S}^1, \dots, \mathcal{S}^B)$ for each pooling level:
\begin{equation}
    \mathcal{S} = se(\mathbf{S})
\end{equation}
The Static Encoder comprises $B$ consecutive blocks made of Skeletal Linear and Pool operators with a Leaky ReLU activation function. Each static learned feature $\mathcal{S}^i$ is extracted after the execution of each block $i$ where $0 \leq i < B$. The dynamic decoder later uses these features. 

Secondly, the Dynamic Encoder, $de$, takes as input the displacement $\mathbf{D}$ and a subset $\mathbf{Q^S}$ of $\mathbf{Q}$ containing only the sparse input (hands, head, root and toes joints) to encode the primal skeleton $\mathcal{P}$:
\begin{equation}
    \mathcal{P} = de \Big( \mathbf{D}, \mathbf{Q^S} \Big)
\end{equation}
The Dynamic Encoder uses $B$ consecutive blocks of Skeletal Convolutions (with a stride of two) and Leaky ReLU activation functions. We represent the primal skeleton as in Figure~\ref{fig:skeleton_pooling}; however, it can be thought of as six joints with multiple learned features each. Finally, the decoder $d$ takes the primal skeleton as input and reconstructs the full-body pose with Skeletal Unpooling, Temporal Upsampling, and Skeletal Convolution (with a stride of one) operations:
\begin{equation}
    \mathbf{\hat{Q}^G} = d(\mathcal{S}, \mathcal{P})
\end{equation}
As we execute the skeletal convolutions with a stride of two in the Dynamic Encoder, the temporal dimension is halved after each block. Then, we use the Temporal Upsampling operation to linearly upsample the frames by two, hence, restoring the initial length of the animation. At the execution of each block $i$, $\mathbf{S}^i$ is added to the convolution result to consider the static structure of the skeleton.

We found that enforcing the end-effectors' position directly on the pose synthesized by the generator using Forward Kinematics-based (FK) losses~\cite{Pavllo:2018, Pavllo:2020} made the training process more difficult, unstable, and unpredictable. Instead, by utilizing dual quaternions in root space and the following Mean Squared Error reconstruction loss we obtained the most favorable results: 
\begin{equation}
    \mathcal{L}_{DQ} = MSE \left(\mathbf{\hat{Q}^G}, \mathbf{Q} \right)
\end{equation}

\paragraph{Learned IK}
\begin{figure*}[hbt]
  \includegraphics[width=1\linewidth]{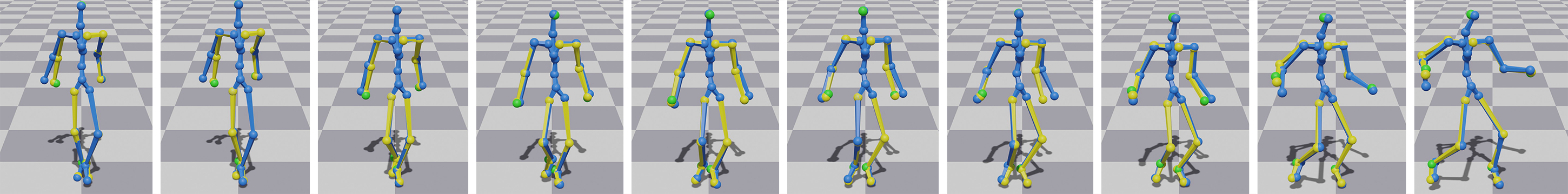}
  \caption{The generator is often not able to accurately match the leaf joints with the end-effectors, which is resolved by the learned IK. The poses synthesized by the generator are shown in blue, those corrected by the learned IK in yellow, while the sparse input data are in green.}
  \label{fig:gt_lik}
\end{figure*}
\begin{figure*}[hbt]
  \includegraphics[width=1\linewidth]{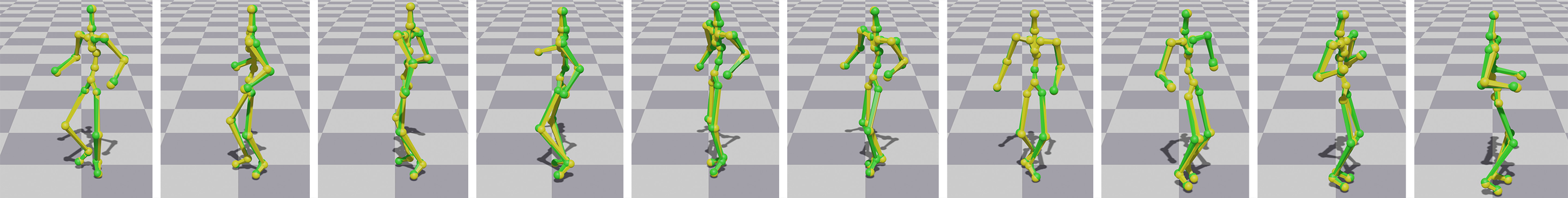}
  \caption{Motion generated by our full approach (yellow) compared to ground truth (green).}
  \label{fig:poses}
\end{figure*}
The generator synthesizes high-quality and continuous human poses. 
However, for certain use cases such as VR, precise positioning of end-effectors may be necessary \citep{Haoran:2023}. We found that the generator's convolutional-based architecture struggles to preserve actual positions and rotations from the limited input data $\mathbf{Q^S}$, resulting in inaccuracies when positioning the end-effectors even when FK-based losses \citep{Pavllo:2018, Pavllo:2020} are utilized, as shown in Section~\ref{sec:ablation}. To address this, we train a series of feedforward neural networks, each specialized in a particular body limb, to make slight adjustments to the limb's pose. Figure~\ref{fig:gt_lik} shows the differences between the pose synthesized by the generator before and after the learned IK stage. We employ IK networks only for the arms and legs; we do not have a network for the head end-effector as our skeleton only has two joints (neck and head) and the generator already produces satisfactory results. 

Each network takes as input the dynamic and static components and the end-effector translation and rotation of the corresponding limb. As a result, each network returns the modified pose for its corresponding limb, for example, the left arm. All adjusted poses are then combined and override the results given by the generator. It is important to note that the learned IK solver never overrides the spine. The results of the full approach is shown in  Figure~\ref{fig:poses}. We add two losses $\mathcal{L}_{S}$ and $\mathcal{L}_{Reg}$. The first loss guarantees precise positioning of end-effectors, while the second ensures that the pose generated by the generator is upheld.
$\mathcal{L}_{S}$ uses FK to compare the positions and rotations with those of the end-effectors:
\begin{equation}
    \mathcal{L}_{S} = MSE \left( FK(\mathbf{Q}), FK(\mathbf{\hat{Q}^{IK}}) \right)
\end{equation}
where $\mathbf{\hat{Q}^{IK}}$ is the final pose after the execution of the learned IK networks. Note that this loss is only computed over the joints related to the end-effectors of the limbs, i.e., hands and toes. When two or more end-effectors are not within reaching bounds, the optimization policy is implicitly learned by the generator. Subsequently, as the learned IK operates on each limb independently, it makes adjustments to each limb based on the output of the generator.

Next, we use a regularization loss that enforces the final pose to be as close as possible to the one synthesized by the generator. This loss is necessary because the learned IK subnetworks are unaware of the full-body pose, and, thus may create unrealistic poses. Moreover, it cannot guarantee continuity since it has no access to previous poses. Thus, $\mathcal{L}_{Reg}$ is needed to allow for minor adjustments while maintaining the pose created by the generator:
\begin{equation}
    \mathcal{L}_{Reg} = MSE \left( FK(\mathbf{\hat{Q}^G}), FK(\mathbf{\hat{Q}^{IK}}) \right)
\end{equation}

The final loss used to train the learned IK is a weighted combination $\mathcal{L}_{S} + \lambda \mathcal{L}_{Reg}$ to control the tradeoff between end-effector accuracy and pose quality. In our experiments we used $\lambda = 0.1$. Note also that $\mathcal{L}_{S}$ and $\mathcal{L}_{Reg}$ are not computed over the same joints, $\mathcal{L}_{S}$ is computed for the end-effectors and $\mathcal{L}_{Reg}$  for all non-end-effectors joints.

Our system estimates dual quaternions for all joints, thus estimating both translation and rotation. Although dual quaternions aid the network in understanding motion, when we animate the characters we preserve the original skeleton offsets used for computing $\mathcal{L}_{S}$ (not predicted ones).

\subsection{Network Training}
\label{subsection: Network Training}
We implemented our system in PyTorch \citep{PyTorch:2019} using the AdamW optimizer \citep{Loshchilov:2019}, with a batch size of 256 and a learning rate of $10^{-4}$. For training, we used our own motion capture database with approximately one million poses at 60 frames per second (${\sim}4.5$ hours) and 9 different actors. Users were captured using an Xsens Awinda motion capture system while performing a series of activities such as locomotion, warm-up and workout exercises, sitting, playing VR games, and dancing. We ensured that right/left limbs are equally represented by mirroring the animation sequences in the horizontal axis, thus resulting in two million poses (${\sim}9$ hours). During training, each motion sequence was split in windows of 64 frames with a stride of eight frames. All components, both the generator and learned IK, are trained at the same time in an end-to-end fashion. At each training iteration, we optimize the parameters of the generator using the loss $L_{DQ}$ and then optimize the parameters of the learned IK while freezing the parameters of the generator. For evaluation, we retrained our system with the DanceDB~\citep{DanceDB:2019} as explained in Section~\ref{sec:comparison}. The training took around 13 hours for our database, and 6 hours for the DanceDB, on a PC equipped with an Intel Core i7-12700k CPU, 32GB of RAM and an NVIDIA GeForce RTX 3090 GPU.
\section{Virtual Reality Controller}
\label{Section:VR}

\begin{figure}[t]
  \includegraphics[height=0.55\linewidth]{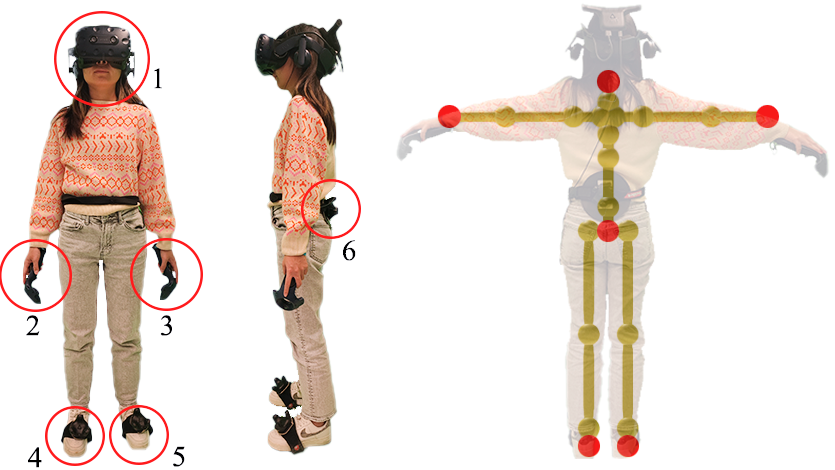}
  \caption{(Left) Sensor placement in a virtual reality setting: (1) head-mounted display; (2 and 3) hand-held controllers; (4 and 5) foot trackers; (6) pelvis or root tracker. (Right) Joints used to compute the offsets with the sensors.}
  \label{fig:sensors}
\end{figure}

We used our system to animate a full-body avatar in VR from a sparse set of sensors providing positional and rotational information. Specifically, we used a HTC VIVE Pro Head-Mounted Display (HMD) with two hand-held controllers and three HTC VIVE Trackers placed on the feet and back (at hip level) as shown in Figure~\ref{fig:sensors}. These sensors require at least one base station (laser projector) to track positional and rotational information.

\begin{figure}[tbh]
  \includegraphics[width=1.0\linewidth]{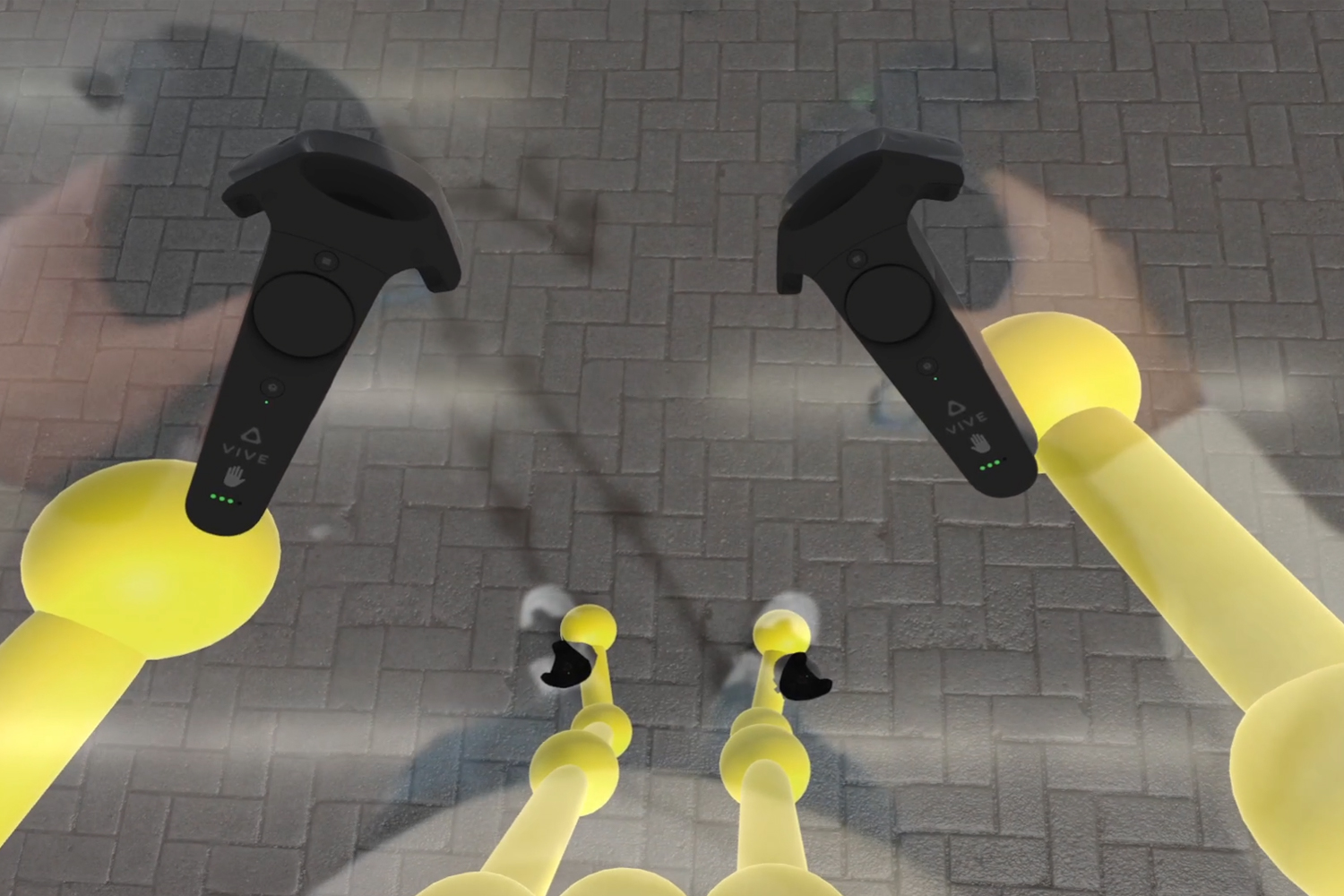}
  \caption{A virtual skeleton rendered over the user as captured by the HMD camera. Note that the visible offsets between physical and virtual controllers are due to the pass-through mode distortion.}
  \label{fig:firstperson}
\end{figure}

Recent studies~\citep{Winkler:2022} have simulated sensor information with respect to the body joints so that the neural network can learn to generate poses. However, we noticed that there is considerable variability in how users hold hand-held controllers and place the trackers, which could lead to calibration difficulties when assuming a fixed sensor placement. 
Instead, our architecture gets as input the positions of the joints directly. Then, similar to the Walk-In-Avatar approach~\citep{Ponton:2022}, we have a calibration step at the beginning of the application in which the avatar appears in T-Pose, and we ask the user to enter the avatar and position themselves inside (see Figure~\ref{fig:sensors}). We assume that user dimensions are known so that bone lengths can be resized to match the user. Some user dimensions can be automatically computed from the sparse input while others are manually introduced, similarly to~\citet{Ponton:2022}. When ready, the user presses any button and our method calculates the offsets between the sensors and their related joints. We use this information, combined with the sensor's positions and angles, as input to our network. 
An avatar animated in VR with SparsePoser is shown in Figure~\ref{fig:firstperson}; note that the visible offsets between physical and virtual controllers are due to the pass-through mode distortion.

At run-time, we keep track of the last $T-1$ frames (in our experiments we use $T = 64$), which, together with the current frame, we use to construct $\mathbf{Q} \in \mathbb{R}^{T \times J \times 8}$. We avoid using future frames when implementing our system for virtual reality to minimize latency, which is crucial to maintain immersion. However, as shown in Section~\ref{sec:comparison}, incorporating access to future information into the pose prediction process can improve the quality of the pose, which may be required for certain applications such as motion capture.
The displacement $\mathbf{D} \in \mathbb{R}^{T \times 3}$ is extracted by the difference in positions of the root sensor. Finally, the static component $\mathbf{S} \in \mathbb{R}^{J \times 3}$ is retrieved directly from a skeleton with the user's dimensions. The output of the static encoder can be fixed for a given subject.

After the generator is executed, it outputs a list of poses (of length $T$), of which only the last one is provided to the learned IK part. The final pose adjusted by the learned IK is used to animate the VR avatar. Finally, we position the avatar in the virtual world using the root sensor position plus the offset computed during the Walk-In-Avatar step. At first, we attempted to predict the movement of the root directly from the network but we encountered issues with positional drift and sliding of the foot. As a result, we opted to enforce the root sensor position and let the network adjust the pose accordingly.

\section{Experiments and Evaluation}
\label{Section:Evaluation}

In this section, we compare our method with previous work, perform an ablation study to evaluate the main components, and assess the use of the system with different user dimensions, i.e., user height and proportions. We performed all the evaluations in real time, exactly mimicking real-world use.

\subsection{Comparison} \label{sec:comparison}
To the best of our knowledge, there are no data-driven methods for reconstructing full-body poses from a sparse set of sensors providing positional and rotational information. Nonetheless, there are some methods able to synthesize plausible poses from three 6-DoF sensors (HMD and two hand-held controllers). The state-of-the-art method is AvatarPoser (AP)~\citep{Jiang:2022b} which employs a Transformer model to generate full-body poses and uses an optimization-based IK method to refine the arms. We extended the implementation of AP to work with six 6-DoF sensors to enable a fair comparison with our approach. We will refer to our extended implementation of AP as the Extended AvatarPoser (EAP). Specifically, we modified the input layer of the Transformer model while maintaining the training procedure and the remainder of the code.

We also evaluate our method against Final IK (FIK)~\citep{FinalIK}, which is a state-of-the-art IK method for animating full-body VR avatars when using a sparse set of 6 degrees of freedom (DoF) trackers. Finally, we compare with other state-of-the-art data-driven methods that reconstruct full-body poses from IMU sensors, such as: TransPose (TP)~\citep{Yi:2021}, and Physical Inertial Poser (PIP)~\citep{Yi:2022}. 
Although comparisons with AP and FIK enable us to evaluate the quality of our method with 6-DoF sensors, it is essential to compare with IMU-based methods to gain a comprehensive understanding of our approach. This is because the use of 6-DoF sensors does not necessarily ensure superiority over IMU-based methods. Additionally, comparing with the wider body of literature on full-body reconstruction provides a broader context for assessing our overall performance gains.

As the generator is convolution-based, we use a window of 64 frames for real-time predictions. When predicting a new pose, we fill this window with past frames of sparse data, the current data, and, optionally, future data. When latency is not an issue, e.g., to generate poses offline from an already captured sequence, we can allow the system to have access to some future information to improve quality. Our system, labeled as \emph{Ours-7} in Table~\ref{tab:comparison}, uses a window of 64 frames, including 56 past frames, the current frame, and 7 future frames. Similarly, \emph{Ours-0} uses 63 past frames, the current frame, but no future frames, resulting in no added latency. In comparison, TransPose uses 5 future frames, while AP, Final IK, and PIP do not use future information.

We conduct a qualitative and quantitative evaluation of our method against EAP, AP, Final IK, TransPose, and PIP. Please refer to the supplementary video for an animated version of our results.

\paragraph{Qualitative} 

\begin{figure}[t]
  \includegraphics[width=1.0\linewidth]{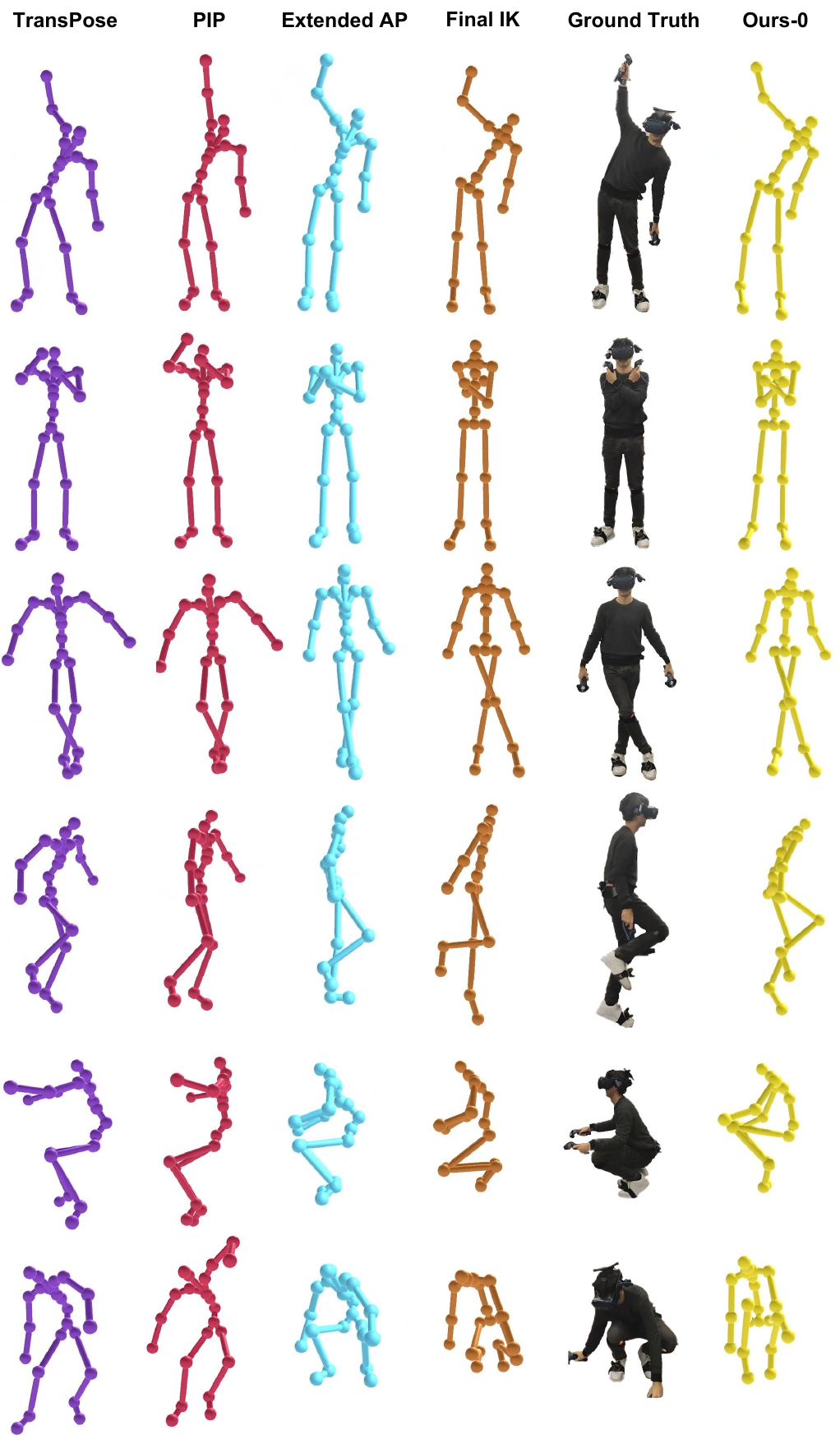}
  \caption{Qualitative comparisons between TransPose, PIP, Extended AvatarPoser, FinalIK, and our method with no added latency (Ours-0). Poses were recorded using HTC VIVE and six IMUs from the Xsens Awinda mocap.}
  \label{fig:comparison}
\end{figure}

In order to provide a visual comparison of our method with related work, selected frames from the video are shown in Figure~\ref{fig:comparison}. In this experiment, we simultaneously collected positional, rotational, and raw IMU data (accelerations and orientations) using the HTC VIVE system and six IMUs from the Xsens Awinda motion capture system. To make it easier to visually compare the poses, the root is fixed in the generated poses.

Both TransPose and PIP generate natural human-like poses in most cases, however, they face challenges when dealing with poses that involve a certain level of ambiguity from the sparse input; for example, when the user crosses two end-effectors, such as hands or feet, or when the user is crouching or lying on the ground. Overall, the movement reconstructed by these methods is often overly smoothed and fails to precisely position the end-effectors. In contrast, Final IK is able to precisely match the end-effectors but fails to reconstruct the real orientations of the joints. For instance, as seen in the fourth row of Figure~\ref{fig:comparison}, the position of the right foot is correct, but the lower leg appears parallel to the ground, differing from the ground truth. In addition, poses often appear too stiff and robotic. Extended AvatarPoser performance lies within an intermediate range, as it generates natural-looking poses in most scenarios. However, its limitations become apparent when it fails to accurately position end-effectors in some instances, resulting in a smoothed pose. This is particularly evident in situations where the pose is ambiguous, as demonstrated in the third row of Figure~\ref{fig:comparison}. Our method, in contrast, is able to position the end-effectors accurately, similar to Final IK, while also maintaining the natural appearance of the poses and correctly matching the joint rotations when compared to the ground truth. We believe our method produces more accurate results due to the two-stage approach, which combines the strengths of a convolutional-based pose generator and a learned IK for accurate positioning. 

\begin{table*}[t]
\footnotesize
\caption{Real time evaluation on HUMAN4D~\citep{HUMAN4D:2020} and SOMA~\citep{SOMA:2021}. We train our method with the DanceDB~\citep{DanceDB:2019} and evaluate it with no added latency (\emph{Ours-0}) and with access to 7 future frames (\emph{Ours-7}). We compare it with state-of-the-art methods using IMU sparse input (TransPose (TP)~\citep{Yi:2021} and Physical Inertial Poser (PIP)~\citep{Yi:2022}) and 6-DoF trackers (Extended AvatarPoser (EAP) from the original paper~\citep{Jiang:2022b}, Final IK (FIK)~\citep{FinalIK}). The table shows the mean and standard deviation (in parenthesis) of the \emph{Positional Error} (Pos), \emph{Rotational Error} (Rot), \emph{End Effector Positional Error} (EE Pos), \emph{Root Error} (Root), \emph{Jitter}, and \emph{Velocity Error} (Vel).}
\label{tab:comparison}
\begin{center}
\begin{tabular*}{\textwidth}{@{\extracolsep{\fill}}l *{2}{c c c c c c}}
  \hline
   & \multicolumn{6}{c}{HUMAN4D} & \multicolumn{6}{c}{SOMA} \\
   \cline{2-7} \cline{8-13}
   & \multicolumn{2}{c}{Pose Quality} & \multicolumn{2}{c}{EE Accuracy} & \multicolumn{2}{c}{Smoothness} & \multicolumn{2}{c}{Pose Quality} & \multicolumn{2}{c}{EE Accuracy} & \multicolumn{2}{c}{Smoothness} \\
   \cline{2-3} \cline{4-5} \cline{6-7} \cline{8-9} \cline{10-11} \cline{12-13}
            & Pos & Rot & EE Pos & Root & Jitter & Vel
            & Pos & Rot & EE Pos & Root & Jitter & Vel\\ 
            & ($\scriptscriptstyle cm$) & ($\scriptscriptstyle deg$) & ($\scriptscriptstyle cm$) & ($\scriptscriptstyle cm$) & ($\scriptscriptstyle 10^2m/s^3$) & ($\scriptscriptstyle cm/s$)
            & ($\scriptscriptstyle cm$) & ($\scriptscriptstyle deg$) & ($\scriptscriptstyle cm$) & ($\scriptscriptstyle cm$) & ($\scriptscriptstyle 10^2m/s^3$) & ($\scriptscriptstyle cm/s$)\\
  \hline
  EAP        & 5.34(5.06) & 10.3(9.21) & 7.82(5.30) & \textbf{0.00(0.00)} & 5.45(18.9) & 11.0(23.6)
            & 5.64(5.23) & 9.31(8.02) & 8.86(5.41) & \textbf{0.00(0.00)} & 5.57(18.9) & 12.5(24.5)\\
  FIK       & 3.62(5.22) & 11.7(19.2) & \textbf{1.11(1.37)} & 1.03(0.98) & 2.99(15.0) & \textbf{5.91(21.4)}
            & 3.01(4.07) & 11.4(19.3) & \textbf{1.22(1.07)} & 1.36(0.68) & 3.49(15.2) & \textbf{6.45(21.0)}\\
  TP        & 6.08(5.96) & 7.19(8.33) & 8.77(6.05) & 27.5(17.8) & 3.70(6.16) & 17.4(26.3)
            & 6.14(6.12) & 6.82(7.42) & 9.04(6.03) & 38.0(22.7) & 4.17(7.33) & 19.2(27.3)\\
  PIP       & 6.61(6.73) & 7.94(8.91) & 9.62(6.84) & 22.1(10.8) & \textbf{1.19(3.76)} & 12.7(20.0)
            & 6.06(5.85) & 7.09(7.53) & 9.10(5.88) & 33.1(18.2) & \textbf{1.34(4.41}) & 13.9(19.9)\\
  Ours-0    & 2.90(2.75) & 5.84(6.00) & 3.51(2.64) & \textbf{0.00(0.00)} & 3.30(5.37) & 9.93(13.8)
            & 2.72(2.57) & 5.72(5.70) & 3.36(2.42) & \textbf{0.00(0.00)} & 3.99(7.34) & 11.8(16.2)\\
  Ours-7    & \textbf{2.49(2.40)} & \textbf{4.98(5.04}) & 2.81(2.12) & \textbf{0.00(0.00)} & 2.83(5.05) & 7.00(9.97)
            & \textbf{2.22(2.13)} & \textbf{4.62(4.64)} & 2.49(1.82) & \textbf{0.00(0.00)} & 3.39(7.00) & 8.28(11.6)\\
  \hline
\end{tabular*}
\end{center}
\end{table*}

\begin{table*}[t]
\footnotesize
\caption{Real time evaluation on HUMAN4D~\citep{HUMAN4D:2020} and SOMA~\citep{SOMA:2021}. The table compares the same metrics as  Table~\ref{tab:comparison}, but only considers upper-body joints to ensure a fair comparison with original AvatarPoser implementation (AP), which uses only three sensors.}
\label{tab:comparison:upperbody}
\begin{center}
\begin{tabular*}{\textwidth}{@{\extracolsep{\fill}}l *{2}{c c c c c}}
  \hline
   & \multicolumn{5}{c}{HUMAN4D} & \multicolumn{5}{c}{SOMA} \\
   \cline{2-6} \cline{7-11}
   & \multicolumn{2}{c}{Pose Quality} & EE Accuracy & \multicolumn{2}{c}{Smoothness} & \multicolumn{2}{c}{Pose Quality} & \multicolumn{1}{c}{EE Accuracy} & \multicolumn{2}{c}{Smoothness} \\
   \cline{2-3} \cline{4-4} \cline{5-6} \cline{7-8} \cline{9-9} \cline{10-11}
            & Pos & Rot & EE Pos & Jitter & Vel
            & Pos & Rot & EE Pos & Jitter & Vel\\ 
            & ($\scriptscriptstyle cm$) & ($\scriptscriptstyle deg$) & ($\scriptscriptstyle cm$) & ($\scriptscriptstyle 10^2m/s^3$) & ($\scriptscriptstyle cm/s$)
            & ($\scriptscriptstyle cm$) & ($\scriptscriptstyle deg$) & ($\scriptscriptstyle cm$) & ($\scriptscriptstyle 10^2m/s^3$) & ($\scriptscriptstyle cm/s$)\\
  \hline
  AP    & 4.15(3.57) & 10.0(9.07) & 6.63(3.59) & 3.58(9.89) & 8.17(13.9)
                & 3.98(3.59) & 8.10(7.04) & 6.68(3.37) & 3.75(10.4) & 8.79(14.1)\\
  Ours-0& 3.09(2.81) & 6.53(6.54) & 3.96(2.67) & 2.57(4.03) & 8.96(12.2)
                & 2.83(2.60) & 6.29(6.17) & 3.77(2.46) & 2.98(5.19) & 10.1(13.2)\\
  Ours-7& \textbf{2.71(2.45)} & \textbf{5.59(5.52)} & \textbf{3.24(2.13)} & \textbf{2.11(3.68)} & \textbf{6.31(8.82)}
                & \textbf{2.38(2.18)} & \textbf{5.13(5.08)} & \textbf{2.90(1.86)} & \textbf{2.40(4.76)} & \textbf{7.12(9.46)} \\
  \hline
\end{tabular*}
\end{center}
\end{table*}

\paragraph{Quantitative}
We test our method using two datasets from AMASS \citep{AMASS:2019} that have not been used for training in the learning-based methods:
HUMAN4D~\citep{HUMAN4D:2020} and SOMA~\citep{SOMA:2021}, which contain a variety of human activities captured by commercial marker-based motion capture systems. We chose AMASS as it is a well-known human motion database and is compatible with SMPL~\citep{SMPL:2015}, which is required by the code provided by the authors of AvatarPoser, TransPose, and PIP. In line with previous works that have trained their networks using multiple datasets from AMASS, our system is trained using DanceDB~\citep{DanceDB:2019}, which is also part of AMASS. We also retrained AvatarPoser with the DanceDB. Because our approach relies on joint information as input, there is no need to synthesize VR trackers. Instead, we directly use the orientations from the databases and apply Forward Kinematics to obtain the positions of the end-effectors.

Similar to previous work \citep{Jiang:2022b, Yi:2021, Yi:2022, Jiang:2022}, we evaluate the performance of our method using different metrics:
\begin{itemize}
    \item \emph{Positional Error} (Pos) measures the mean Euclidean distance error of all joints in centimeters. The root position is aligned with the ground truth data.
    \item \emph{Rotational Error} (Rot) measures the mean global rotation error of all joints in degrees. We compute the distance between two rotations represented by rotation matrices $R_0$ and $R_1$ as the angle of the difference rotation represented by the rotation matrix $D = R_0 R_1^T$.
    \item \emph{End Effector Positional Error} (EE Pos) measures the mean Euclidean distance error of end-effectors (excluding the root) in centimeters. The root position is aligned with the ground truth data.
    \item \emph{Root Error} (Root) measures the mean Euclidean distance error of the root joint in centimeters.
    \item \emph{Jitter} measures the mean jerk of all joints in ten squared meters per second cubed. Jerk is the third derivative of position with respect to time, i.e., the rate of change of the acceleration \citep{Flash:1985}. We use it as a measure of the smoothness of the motion.
    \item \emph{Velocity Error} (Vel) measures the mean velocity error of all joints in centimeters per second. The velocity is computed by forward finite differences.
\end{itemize}

We group these metrics into three main categories: \emph{pose quality}, \emph{end-effector accuracy}, and \emph{smoothness}. 
To evaluate the overall pose quality of the generated data, we use the \emph{Positional Error} and \emph{Rotational Error} that measure the joint positions and rotations accuracy, respectively, when the root is aligned with the ground truth data. To evaluate end-effector accuracy, we distinguish between the character's placement in the world (\emph{Root Error}) and the positions of the remaining end-effectors (such as the head, hands, and toes) when the root position is aligned. Lastly, motion smoothness is assessed using \emph{Jitter} and \emph{Velocity Error}.

Table~\ref{tab:comparison} presents the comparison results. The goal of our proposed method is to achieve optimal pose quality while also maximizing end-effector accuracy. Reconstructing full-body poses from sparse data is an under-constrained problem, therefore, a balance must be struck between the two metrics to achieve optimal results. Our method balances the competing demands of high-quality poses and accurate end-effector positioning without negatively impacting the overall human-like appearance of the pose.

It can be observed that Final IK, being an inverse kinematics method, effectively tracks the end-effectors but struggles in synthesizing natural poses, and often introduces jittering artifacts with abrupt changes in direction.
Conversely, methods such as TransPose and PIP, since they use IMU sensors, can achieve high overall pose quality, but they introduce \emph{Positional Error} and low end-effector accuracy. Our model achieves the highest scores for pose quality, regardless of whether future frames are used or not. Additionally, our method greatly improves the accuracy of end-effectors when compared to other data-driven methods, achieving results similar to Final IK, which is specifically designed to minimize the distance between end-effectors and the target. Furthermore, our model outperforms other methods in \emph{Root Error} as we do not predict the root position, but constrain it based on the root sensor and let the networks adjust the pose. This aspect is crucial for self-avatar animation as it keeps the user correctly positioned with the virtual avatar. In terms of smoothness, PIP has the best results in \emph{Jitter} but the worst in \emph{End-Effector Positional Error}, which suggests that they are missing the high-frequency details of the movement. In contrast, our method provides a good balance as it obtains the second-best scores in \emph{Jitter} and \emph{Velocity Error} while maintaining high end-effector accuracy with a smaller variance. This suggests fewer large changes in pose between frames and fewer jittering artifacts, resulting in less noticeable popping artifacts in the animation. 

Finally, our method outperforms Extended AvatarPoser across all metrics (except for \emph{Root Error}, since both methods introduce no root error). We consider AvatarPoser as our baseline since it also uses 6-DoF trackers, but employs the well-established Transformer architecture. Hence, the performance of our approach is not solely attributable to the use of 6-DoF trackers. As we extended the input of AvatarPoser's Transformer model to include six 6-DoF trackers instead of the original three, to further validate our findings, we also present in Table~\ref{tab:comparison:upperbody} a comparison of the same metrics but only for the upper-body joints synthesized with the original AvatarPoser implementation. Remarkably, even when focusing solely on the upper-body joints, our approach still clearly outperforms AvatarPoser.

We attribute the superior performance of our approach compared to the Extended Avatar Poser to the specialized architectural composition of our networks. As opposed to Transformers, originally crafted for natural language processing, our method deploys skeleton-aware operations intrinsically designed to accommodate the hierarchical structure of the human skeleton. In addition, our dual-stage strategy employs a time-aware network using convolutions, enabling them to learn a comprehensive representation of human motion, at the expense of losing some high-frequency motion details. Nonetheless, our method can recover the high-frequency details through the utilization of the learned IK. Crucially, we posit that our learned IK, trained in an end-to-end fashion with the generator, is capable of learning an optimization policy that more accurately replicates natural human motion, surpassing the traditional optimization-based IK employed in AvatarPoser.

\subsection{Ablation Study} 
\label{sec:ablation}
As outlined in the previous section, our goal is to achieve both optimal pose quality and maximum end-effector accuracy. In this section, we describe an ablation study to examine the impact of each of the components of our network on the balance between pose quality and end-effector accuracy. We trained and evaluated our system on the same datasets as in Section~\ref{sec:comparison}. For a fair comparison, all experiments in this section had access to the 7 future frames, matching the conditions of the \emph{Ours-7} version, which all ablation tests are compared against. All results are listed in Table~\ref{tab:ablation}; please refer to the supplementary materials for an animated version of these results.

In the initial experiment, we assess the effect of using the generator alone, without the learned IK. We compared two versions: first (No Learned IK in Table~\ref{tab:ablation}), the learned IK is not used and the rest of the pipeline remains intact; second (Generator $\mathcal{L}_{S}$ in Table~\ref{tab:ablation}), a Forward Kinematics loss similar to $\mathcal{L}_{S}$ was added to compare the pose generated by the generator and the ground truth, $MSE \left( FK(\mathbf{Q}), FK(\mathbf{\hat{Q}^{G}}) \right)$.

In this case, the only metric that showed improvement was jitter. However, it was observed that the reconstructed motion failed to maintain high-frequency details, resulting in lower performance in other metrics. In the second case, when the FK-based loss is added to the output of the generator, we observed a slight decrease in rotational error, but a notable increase in both end-effector positional error and overall positional error when compared to the case of using the learned IK component. Thus, these findings suggest that the inclusion of the learned IK component significantly improves the end-effector accuracy while preserving the high-quality poses synthesized by the generator. It is worth noting that, by improving the end-effector positions and maintaining a low rotational error, the overall positional error is decreased as the limbs are correctly positioned.

Since the learned IK operates on each limb independently, it lacks the ability to take into account the overall body pose. Therefore, when omitting the $\mathcal{L}_{Reg}$ loss term (No $\mathcal{L}_{Reg}$ in Table~\ref{tab:ablation}), while there may be a slight improvement in end-effector accuracy, a significant decline in pose quality is observed. By looking at the generated poses, it can be seen how the limbs are attempting to reach the end-effectors at the cost of synthesizing non-human-like motion. As such, the inclusion of the $\mathcal{L}_{Reg}$ loss term leverages the strengths of the generator with the learned IK, resulting in improved pose quality and end-effector accuracy.

Additionally, to evaluate the impact of the skeletal-aware operations, we define a baseline method (No Skeletal Op. in Table~\ref{tab:ablation}). Specifically, we replaced the previously-used skeletal convolutions with conventional one-dimensional convolutions and modified the skeletal unpooling to allow unpooled joints to receive information from all joints instead of just neighboring ones. Not accounting for the joint adjacency resulted in a significant decline in performance across all metrics. By inspecting the visual results, we believe that allowing convolutions to have access to all joints produces an average effect that results in an
overly smooth motion.
\begin{table*}
\footnotesize
\caption{Ablation study on HUMAN4D and SOMA datasets. We trained our method with DanceDB in all experiments. The first two experiments (No Learned IK; Generator $\mathcal{L}_{S}$) do not incorporate the learned IK component, but the second adds a FK-based loss to the output of the generator. The third experiment (No $\mathcal{L}_{Reg}$) removes the $\mathcal{L}_{Reg}$ loss term. The last experiment defines the baseline method (No Skeletal Op.), we replace all skeleton-aware operations with standard one-dimensional convolutions.}
\label{tab:ablation}
\begin{center}
\begin{tabular*}{\textwidth}{@{\extracolsep{\fill}}l *{3}{c c c c c}}
  \hline
   &  \multicolumn{5}{c}{HUMAN4D} & \multicolumn{5}{c}{SOMA} \\
   \cline{2-6} \cline{7-11} 
   & \multicolumn{2}{c}{Pose Quality} & \multicolumn{1}{c}{EE Accuracy} & \multicolumn{2}{c}{Smoothness} & \multicolumn{2}{c}{Pose Quality} & \multicolumn{1}{c}{EE Accuracy} & \multicolumn{2}{c}{Smoothness} \\
   \cline{2-3} \cline{4-4} \cline{5-6} \cline{7-8} \cline{9-9} \cline{10-11}
            & Pos & Rot & EE Pos & Jitter & Vel
            & Pos & Rot & EE Pos & Jitter & Vel\\
            & ($\scriptscriptstyle cm$) & ($\scriptscriptstyle deg$) & ($\scriptscriptstyle cm$) & ($\scriptscriptstyle 10^2cm/s^3$) & ($\scriptscriptstyle cm/s$)
            & ($\scriptscriptstyle cm$) & ($\scriptscriptstyle deg$) & ($\scriptscriptstyle cm$) & ($\scriptscriptstyle 10^2cm/s^3$) & ($\scriptscriptstyle cm/s$)\\
  \hline
  No Learned IK
            & 3.83(3.94) & 5.47(5.56) & 6.39(4.75) & \textbf{1.60(2.85)} & 9.75(15.4)
            & 3.49(3.61) & 5.08(5.15) & 5.83(4.31) & \textbf{2.69(4.00)} & 11.7(18.1)\\
  Generator $\mathcal{L}_{S}$
            & 3.37(3.50) & \textbf{4.73(4.83)} & 5.16(4.06) & 2.14(2.92) & 9.17(14.2)
            & 3.01(3.09) & \textbf{4.29(4.38)} & 4.70(3.51) & 2.70(3.65) & 10.7(16.1) \\
  No $\mathcal{L}_{Reg}$
            & 3.49(3.65) & 9.66(10.3) & \textbf{2.19(1.75)} & 3.04(5.51) & 7.55(11.2)
            & 3.04(3.21) & 8.85(9.69) & \textbf{1.99(1.55)} & 3.53(7.60) & 9.11(13.6)\\
  No Skeletal Op.
            & 14.0(13.7) & 24.1(21.3) & 19.9(16.6) & 6.94(36.4) & 23.9(60.9)
            & 13.2(13.4) & 27.2(24.3) & 20.3(17.1) & 13.4(72.5) & 34.0(99.9)\\
  Ours-7
            & \textbf{2.49(2.40)} & 4.98(5.04) & 2.81(2.12) & 2.83(5.05) & \textbf{7.00(9.97)}
            & \textbf{2.22(2.13)} & 4.62(4.64) & 2.49(1.82) & 3.39(7.00) & \textbf{8.28(11.6)}\\
  \hline
\end{tabular*}
\end{center}
\end{table*}
\begin{figure*}[hbt]
  \includegraphics[width=1\linewidth]{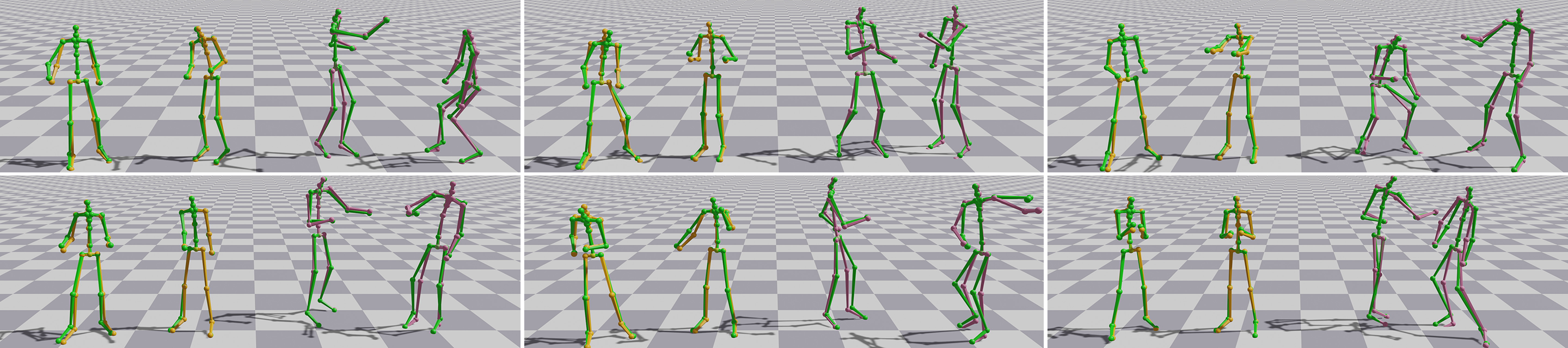}
  \caption{Pose reconstruction for two users of different body proportions (orange and pink, respectively: height: 162~cm and 184~cm; arm span: 151~cm and 187~cm; hip height: 90~cm and 97~cm) while using a VR application.}
  \label{fig:users}
\end{figure*}

\subsection{Pose Representation}
In our method, we use dual quaternions, as proposed by \citet{Andreou:2022}, as our pose representation because they offer a unified and concise representation that includes both rotation and translation information.
Through our experimentation, we have discovered that utilizing dual quaternions results in superior reconstruction of poses and continuity compared to other commonly used pose representations such as quaternions or \emph{ortho6D} \citep{Zhou:2019}. To further evaluate the effectiveness of our method, we have modified it to accept both quaternions and \emph{ortho6D} as pose representations, and conducted comparisons with our dual quaternions implementation. We modified both the input and output of the networks. For a fair comparison, we have also added root-space translation information similar to that encoded in the dual quaternions, but using 3D vectors instead. 

The results show that quaternions and \emph{ortho6D} yield similar outcomes in terms of pose quality (\emph{ortho6D} obtained about 5\% lower in \emph{Rotational Error}). In contrast, the use of dual quaternions leads to approximately ${\sim}50\%$ higher pose quality and ${\sim}60\%$ lower \emph{End Effector Positional Error}. Furthermore, dual quaternions exhibit slightly better results in smoothness.

\subsection{User Dimensions Evaluation}
The ability to adjust poses for different users without requiring retraining of the underlying networks, taking into account factors such as height and body shape, is crucial for motion capture and creating a more personalized experience in virtual reality. Unlike previous approaches that use IMU sensors and rely on a fixed skeleton during training, our method includes a Static Encoder to learn skeletal features, and by using dual quaternions as pose representation, the network can adapt to a wide range of proportions.

To evaluate the effectiveness of our method in capturing the motion of users with different body shapes and sizes, we conducted an experiment using our motion capture dataset collected with an Xsens device. For this purpose, we retrained the system omitting the motion data from two users (about 30 minutes of data per user) with distinct physical characteristics (height: 162\,cm and 184\,cm; arm span: 151\,cm and 187\,cm; hip height: 90\,cm and 97\,cm), and used it to evaluate the accuracy of their predicted poses. Our method was able to accurately reconstruct the motion from both users, with a difference of about ${\sim}20\%$ in \emph{Positional Error}, ${\sim}15\%$ in end-effectors' \emph{Positional Error} and ${\sim}5\%$ in \emph{Rotational Error}. Figure~\ref{fig:users} illustrates the pose reconstruction for two users with different body proportions using our system. However, we expect that these differences will decrease as we include a larger variety of users in the training set, as the current dataset only contains data from seven different users.

\subsection{Limitations}

The limitations of our method include its reliance on the quality of the training dataset. As with previous data-driven techniques, our approach may inadvertently learn from inaccuracies or artifacts in the ground truth data or have difficulty generalizing to sparse input that it has not been sufficiently exposed to, such as uncommon wrist rotations. In addition, SparsePoser works best when the input data is within the range of typical human poses. However, if one of the tracking devices is malfunctioning or the input data does not
correspond to a human skeleton, our method may fail to produce a plausible pose. 

Furthermore, our method demands a very specific setup. The system's functionality could be enhanced by enabling it to work with varying numbers of sensors (e.g., HMD and two hand-held controllers) or degrees of freedom (e.g., sensors providing only positional information), which would increase its applicability across different scenarios. Currently, it is necessary to tailor the user's skeleton to maximize the fidelity of the generated motion. By incorporating simpler high-level attributes like height and width instead of each bone length, the usability of our method could be significantly enhanced.

Another limitation is the focus on the skeleton without considering the user's physical body or surrounding surfaces. As such, it may unintentionally synthesize self-penetrations or similarly unrealistic outcomes. Addressing these considerations would increase the flexibility and real-world adaptability of our method.

\section{Conclusions and future work}
\label{Section:Conclusions}

In this paper, we have presented SparsePoser, a new learning-based architecture to synthesize high-quality human motion from sparse input. Our network generates full-body animations from just six trackers, placed on the pelvis (root) and the five endpoints of the human skeleton (head, hands, and feet). 

The comparisons with competing approaches demonstrate that SparsePoser generates animations whose pose quality clearly outperforms state-of-the-art motion reconstruction methods; as our method provides the lowest positional and rotational errors (lowest error mean and lowest error variance). We have shown that such pose quality does not come at the price of end-effector accuracy. In fact, our method beats non-IK methods in terms of end-effector placement.  

The key components of our approach are a convolution-based generator that synthesizes high-quality animations, and learned IK networks that slightly adjust the generated poses to fit the trackers' positions. The generator is an autoencoder that learns the human motion features from the sparse motion input, ensuring smooth animations. The IK adjustments are carried out by feed-forward neural networks, each one specialized in a particular body limb. 

The ablation study has revealed the individual contribution of the main ingredients of SparsePoser, including the role of skeletal-aware vs. 1D convolutions, the encoding of the pose through dual quaternions instead of ordinary quaternions, the learned-IK adjustment, as well as the different loss functions.   

Since SparsePoser runs in real-time and is able to work with no future frames, it is suitable for those applications (including VR) where low latency is critical. The \emph{Ours-0} version has an end-to-end latency similar to state-of-the-art IK-based approaches, as shown in the supplementary video. Furthermore, the accurate positioning of the end-effectors makes it ideal for applications where the avatars interact with other objects, as well as for VR self-avatars. 

Although we tested SparsePoser on VR hardware, its applications go beyond VR, as some companies have just started to provide standalone low-cost 6-tracker systems (e.g., Sony Mocopi). SparsePoser can be used as a cheap motion capture method for varied applications. 

In future work, we plan to evaluate and possibly extend our architecture to deal with sparse data from a different number of trackers (either fewer trackers for even more widespread use, or more trackers to compete with professional mocap systems). We wish to extend our architecture to cope with noisy inputs (e.g., high-latency input from remote avatars in social VR). Finally, we also plan to explore generative models to handle different sensor configurations. \\

\noindent
\textbf{Code and data.}
The complete source code, trained model, animation databases, and supplementary material used in this paper can be found at
\href{https://upc-virvig.github.io/SparsePoser}{https://upc-virvig.github.io/SparsePoser}.
\begin{acks}
This work has received funding from the European Union’s Horizon 2020 research and innovation programme under the Marie Skłodowska-Curie grant agreement No 860768 (CLIPE project), HORIZON-CL4-2022-HUMAN-01 grant agreement No 101093159 (XR4ED), and from MCIN/AEI/10.13039/501100011033/FEDER, UE (PID2021-122136OB-C21). This project has also received funding from the European Union's Horizon 2020 Research and Innovation Programme under Grant Agreement No 739578 and the Government of the Republic of Cyprus through the Deputy Ministry of Research, Innovation and Digital Policy. Jose Luis Ponton was also funded by the Spanish Ministry of Universities (FPU21/01927).
\end{acks}

\bibliographystyle{ACM-Reference-Format}
\bibliography{References}


\end{document}